\documentclass{aa}
\usepackage{natbib}
\bibpunct{(}{)}{;}{a}{}{,} % to follow the A&A style
\usepackage{graphicx}
\usepackage{txfonts}
\usepackage{color}

\begin{document}

   \title{Comparison of the orbital properties of Jupiter Trojan asteroids and Trojan dust}

   %\subtitle{I. Overviewing the $\kappa$-mechanism}

   \author{Xiaodong Liu 
          \and
          J\"urgen Schmidt
          }

   \institute{Astronomy Research Unit, University of Oulu, Finland\\
              \email{xiaodong.liu@oulu.fi} 
             }

   \date{}

% \abstract{}{}{}{}{} 
% 5 {} token are mandatory
 
  \abstract
{In a previous paper we simulated the orbital evolution of dust particles from the Jupiter Trojan asteroids ejected by the impacts of interplanetary particles, and evaluated their overall configuration in the form of dust arcs. Here we compare the orbital properties of these Trojan dust particles and the Trojan asteroids. Both Trojan asteroids and most of the dust particles are trapped in the Jupiter 1:1 resonance. However, for dust particles, this resonance is modified because of the presence of solar radiation pressure, which reduces the peak value of the semi-major axis distribution. We find also that some particles can be trapped in the Saturn 1:1 resonance and higher order resonances with Jupiter. The distributions of the eccentricity, the longitude of pericenter, and the inclination for Trojans and the dust are compared. For the Trojan asteroids, the peak in the longitude of pericenter distribution is about 60 degrees larger than the longitude of pericenter of Jupiter; in contrast, for Trojan dust this difference is smaller than 60 degrees, and it decreases with decreasing grain size. For the Trojan asteroids and most of the Trojan dust, the Tisserand parameter is distributed in the range of two to three.}


   \maketitle

\section{Introduction} \label{section_introduction}
Dust ejecta are generated when hypervelocity interplanetary micrometeoroids impact the surfaces of atmosphereless bodies \citep{kruger1999detection,horanyi2015permanent, szalay2016impact}. As of July 2016, there are about 4266 known Jupiter Trojan asteroids in the $L_4$ swarm and 2341 in the $L_5$ swarm (JPL (Jet Propulsion Laboratory) Small-Body Database Search Engine). In previous work, we estimated the total cross section of the Trojan asteroids that is exposed to the interplanetary flux, and evaluated the production rate of ejecta particles from the Trojan asteroids \citep{liu2018dust}. We also simulated the subsequent evolution of these ejecta particles after ejection, and derived their steady-state spatial configuration.

The orbital parameter distribution of the Trojan asteroids has received considerable interest \citep{jewitt2000population, yoshida2005size}. The motion of the Trojan dust particles is influenced by non-gravitational perturbation forces (see Section \ref{chap:DynamicalModel}) so that their dynamics differs from that of the Trojan asteroids, which is reflected in their respective orbital properties. In this paper, we compare the orbital properties of the Jupiter Trojan asteroids and Trojan dust, which show a similar orbital behavior, but are also different in many aspects.

Because of their low number density, there is no detection yet of Trojan dust particles although upper limits on their cross section were inferred \citep{2000Icar..145...44K}, and our work is based on numerical and analytical investigations. The \textit{Lucy} spacecraft developed by NASA will be the first mission to conduct reconnaissance of Jupiter's $L_4$ and $L_5$ Trojan asteroids \citep{levison2017lucy}. Because the dust particles are samples from the asteroid surfaces their in-situ compositional analysis by future space missions will allow researchers to directly determine the chemical composition of the asteroids \citep{sternovsky2017surface, Toyota2017density}.

\section{Brief description of the dynamical model} \label{chap:DynamicalModel}
In this paper, the orbital parameters are defined with respect to the Jupiter orbital inertial frame $Ox_\mathrm{oi}y_\mathrm{oi}z_\mathrm{oi}$. The $z_\mathrm{oi}$ axis points to the normal of Jupiter's orbital plane at J2000 epoch, the $x_\mathrm{oi}$ axis is taken to be the intersection of Jupiter's orbital plane with Jupiter's equator plane at J2000 epoch, and the $y_\mathrm{oi}$ axis follows the right-hand rule. The Trojans' orbital elements in the J2000 ecliptic reference frame are obtained from the JPL Small-Body Database Search Engine, and then are transformed to the frame $Ox_\mathrm{oi}y_\mathrm{oi}z_\mathrm{oi}$. 

For the dust particles we analyze simulation results from a previous study \citep{liu2018dust}. Here we briefly describe our model. Compared with asteroids, dust particles are subject to non-gravitational forces including solar radiation pressure, Poynting-Robertson (P-R) drag, solar wind drag, and the Lorentz force due to the interplanetary magnetic field. We adopt silicate as the material for Trojan dust. In our simulations, particles in the size range [0.5, 32] $\mu\mathrm{m}$ are included. For each size, 100 particles from $L_4$ Trojans are simulated until they hit Jupiter or their Jovicentric distances lie outside the region of interest, [0.5, 15] AU. For the details of the dynamical model and numerical simulations, the readers are referred to \citet{liu2018dust} and \citet{liu2016dynamics}.

\section{Distribution of semi-major axis $a$ and orbital resonances}
\begin{figure}
\centering
\noindent\includegraphics[width=9cm]{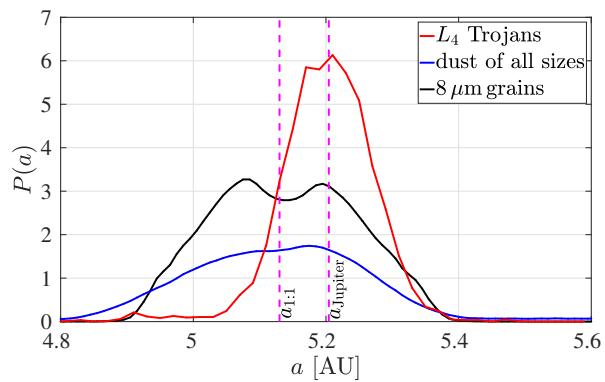}
\caption{\label{L4vsdust_a_distri}Semi-major axis distribution for the $L_4$ asteroids (red) and Trojan dust of all sizes (blue). The black line denotes the semi-major axis distribution for $8 \, \mathrm{\mu m}$ particles. The size-dependent resonant semi-major axis $a_{1:1}$ (Eq.~\ref{eq_a1:1}) is for $8 \, \mathrm{\mu m}$ particles.}
\end{figure}

We start our analysis from the distribution of the semi-major axis for $L_4$ Trojans and Trojan dust (Fig.~\ref{L4vsdust_a_distri}). As expected, the distribution of the Trojans' semi-major axes peaks around $a_\mathrm{Jupiter}$ (5.2 AU), while the peak for the dust population is smaller and much wider. This is mainly because of the effect of solar radiation pressure combined with the Jupiter 1:1 mean motion resonance.

\begin{figure}
\centering
\noindent\includegraphics[width=9cm]{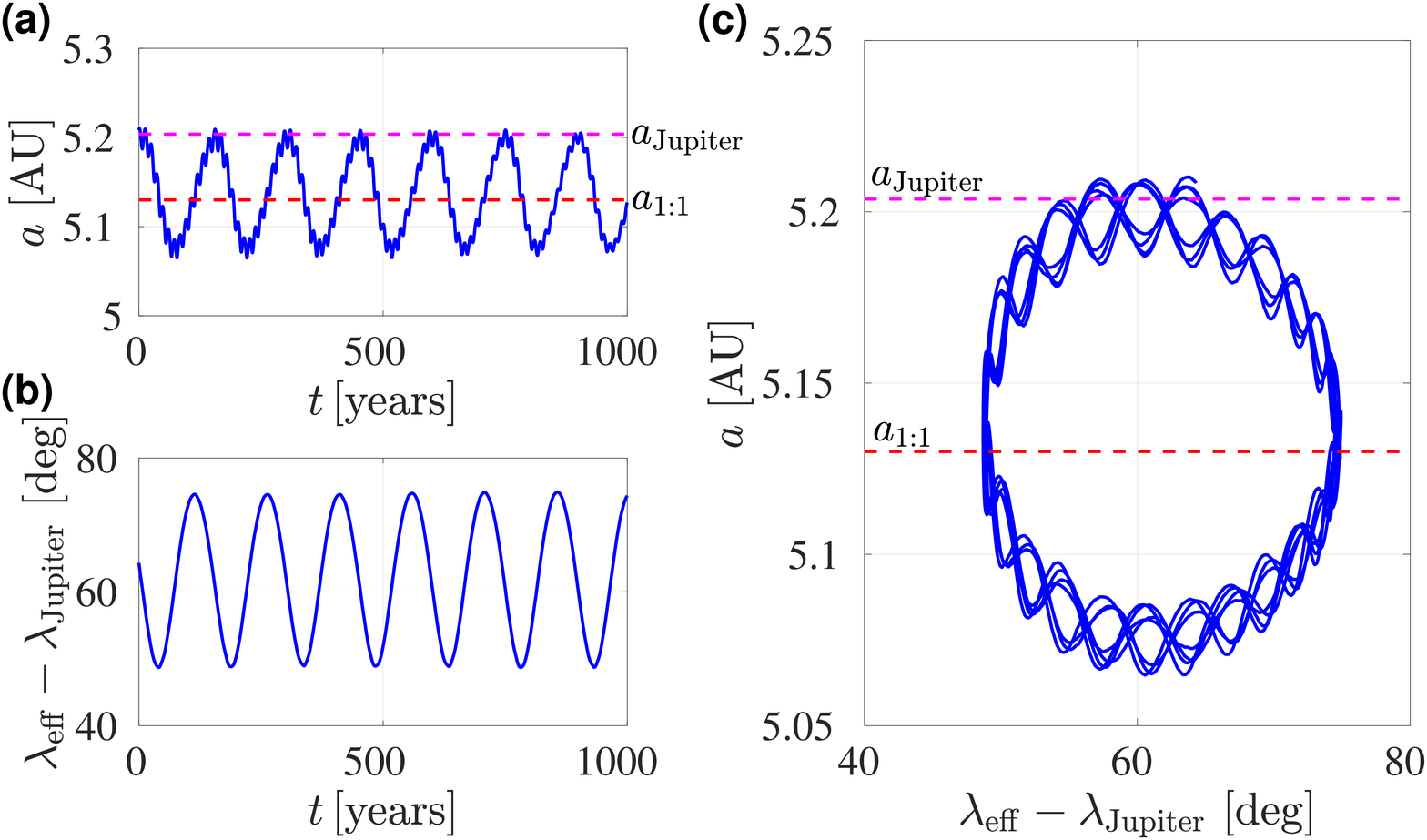}
\caption{\label{Trapped_8um_evolution} Evolution of semi-major axis \textbf{(a)} and resonant argument $\Psi_\mathrm{1:1} = \lambda_\mathrm{eff} -\lambda_\mathrm{Jupiter}$ \textbf{(b)} for an $8 \, \mathrm{\mu m}$ particle. \textbf{(c)}, Semi-major axis versus $\lambda_\mathrm{eff} -\lambda_\mathrm{Jupiter}$ for an $8 \, \mathrm{\mu m}$ particle for 1000 years of its evolution. Here the resonant semi-major axis $a_{1:1}$ is also for $8 \, \mathrm{\mu m}$ particles.}
\end{figure}

The direction of solar radiation pressure is opposite to that of the solar gravitational force. Thus, the effect of solar radiation pressure can be included in terms of the effective solar mass $M_\mathrm{eff}=M_\mathrm{Sun}(1-\beta)$ \citep{kresak1976orbital}, where $M_\mathrm{Sun}$ is the mass of the Sun. Other symbols with subscript ``eff'' in the text are defined when the effective solar mass $M_\mathrm{eff}$ is considered. For grain sizes $\leqslant 2\, \mu\mathrm{m}$, the Trojan particles have short lifetimes \citep{liu2018dust} because the strong solar radiation pressure rapidly expels them away from the $L_4$ region. Larger grains $\geqslant 4\, \mu\mathrm{m}$ can be trapped in the 1:1 resonance with Jupiter for a long time. The effective resonant argument $\Psi_\mathrm{1:1}$ of the particle's 1:1 resonance with Jupiter is expressed as
\begin{equation} \label{eq_psi_eff}
\Psi_\mathrm{1:1} = \lambda_\mathrm{eff} - \lambda_\mathrm{Jupiter} \,,
\end{equation}
where $\lambda$ is the mean longitude. At resonance, $\dot\Psi_\mathrm{1:1}=0$ \citep{murray1999solar}, and the effective semi-major axis at the Jupiter 1:1 resonance is obtained as \citep{liou1995asteroidal,liou1995radiation}
\begin{equation} \label{eq_eff1:1_a}
a_\mathrm{eff,\, 1:1} = a_\mathrm{Jupiter} \left[\frac{M_\mathrm{Sun}(1-\beta)}{M_\mathrm{Sun}+M_\mathrm{Jupiter}}\right]^{1/3} \,.
\end{equation}
Here $M_\mathrm{Jupiter}$ is the mass of Jupiter, and $\beta$ is the ratio of solar radiation pressure relative to solar gravitational force \citep{BURNS:1979wg}. The value of $\beta$ is size-dependent, the dependence of which on grain size for silicate particles can be found in Fig.~4b of \citet{liu2018dust}. \citet{kresak1976orbital} derived the relation between the effective and conventional semi-major axis when dust particles are released at the perihelion and aphelion of a comet. Because of the small eccentricity for the majority of dust particles, this relation can be simplified by neglecting the effect of eccentricity as
\begin{equation} \label{eq_eff&phy_a}
a_\mathrm{eff} = a\frac{1-\beta}{1-2\beta} \,.
\end{equation}
Combining Eqs.~\ref{eq_eff1:1_a} and \ref{eq_eff&phy_a}, the conventional semi-major axis at the Jupiter 1:1 resonance is calculated as
\begin{equation} \label{eq_a1:1}
a_\mathrm{1:1} = a_\mathrm{Jupiter}\frac{1-2\beta}{1-\beta} \left[\frac{M_\mathrm{Sun}(1-\beta)}{M_\mathrm{Sun}+M_\mathrm{Jupiter}}\right]^{1/3} \,.
\end{equation}
It is apparent that $a_\mathrm{1:1}<a_\mathrm{Jupiter}$, and $a_\mathrm{1:1}$ becomes smaller with decreasing grain size because their $\beta$ becomes larger. It follows that solar radiation pressure can efficiently decrease the values of semi-major axes. This explains why the peak in dust density is shifted radially inward from Jupiter (Fig.~7a of \citet{liu2018dust}). As an example, Fig.~\ref{Trapped_8um_evolution} shows the evolution of the semi-major axis and resonant argument for an $8 \, \mathrm{\mu m}$ particle for 1000 years. The dust semi-major axis librates around $a_\mathrm{1:1}$ with nearly constant amplitude. The resonant argument $\Psi_\mathrm{1:1} = \lambda_\mathrm{eff} -\lambda_\mathrm{Jupiter}$ librates around a fixed angle larger than $60^\circ$ (Figs.~\ref{Trapped_8um_evolution}a,b). The small wiggles in Figs.~\ref{Trapped_8um_evolution}a,c are also induced by solar radiation pressure.

The distribution of the semi-major axis for $8 \, \mathrm{\mu m}$ particles is bimodal (Fig.~\ref{L4vsdust_a_distri}). This is because the rate of change for the semi-major axis is slowest near the turning points of the libration (Fig.~\ref{Trapped_8um_evolution}a). As a result, these semi-major axis values contribute more to the distribution. The precise libration amplitudes are different for different particles of the same size, which is the reason why the peaks for $8 \, \mathrm{\mu m}$ particles in Fig.~\ref{L4vsdust_a_distri} are wide. Particles of different grain size ($\geqslant 4\, \mu\mathrm{m}$) also possess this bimodal property. However, the bimodality is buried for the whole Trojan dust population because different peaks of various grain sizes are superimposed onto each other.

\begin{figure}
\centering
\noindent\includegraphics[width=9cm]{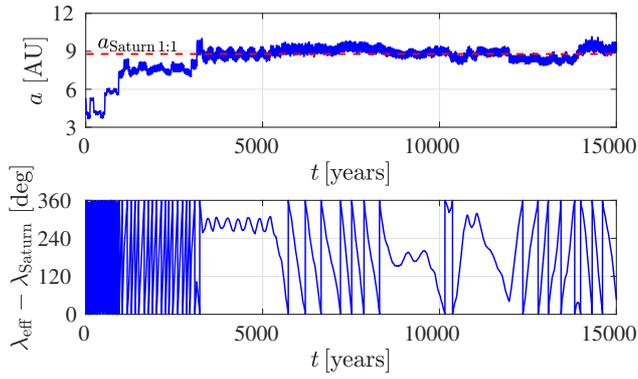}
\caption{\label{Saturn_trapped_2um}Evolution of semi-major axis and resonant argument $\Psi_\mathrm{Saturn \, 1:1} = \lambda_\mathrm{eff} -\lambda_\mathrm{Saturn}$ for a $2 \, \mathrm{\mu m}$ particle. Here $a_\mathrm{Saturn\,1:1}$ is for $2 \, \mathrm{\mu m}$ particles.}
\end{figure}

There exist particles that are not trapped in the Jupiter 1:1 resonance, or which can escape from this resonance. Due to various perturbation forces, these particles are transported either outwards or inwards, which does not occur for the Trojan asteroids themselves. Some grains can be trapped in the Saturn 1:1 resonance. Following the same procedure as for Eqs.~\ref{eq_psi_eff}-\ref{eq_a1:1}, we can calculate the conventional semi-major axis at the Saturn 1:1 resonance. A $2 \, \mathrm{\mu m}$ particle trapped in the Saturn 1:1 resonance is taken as an example (Fig.~\ref{Saturn_trapped_2um}), for which the resonant semi-major axis is $a_\mathrm{Saturn\,1:1} \approx 8.78 \, \mathrm{AU}$. The particle is transported outwards and gets trapped in and escapes the Saturn 1:1 resonance several times.

\begin{figure}
\centering
\noindent\includegraphics[width=9cm]{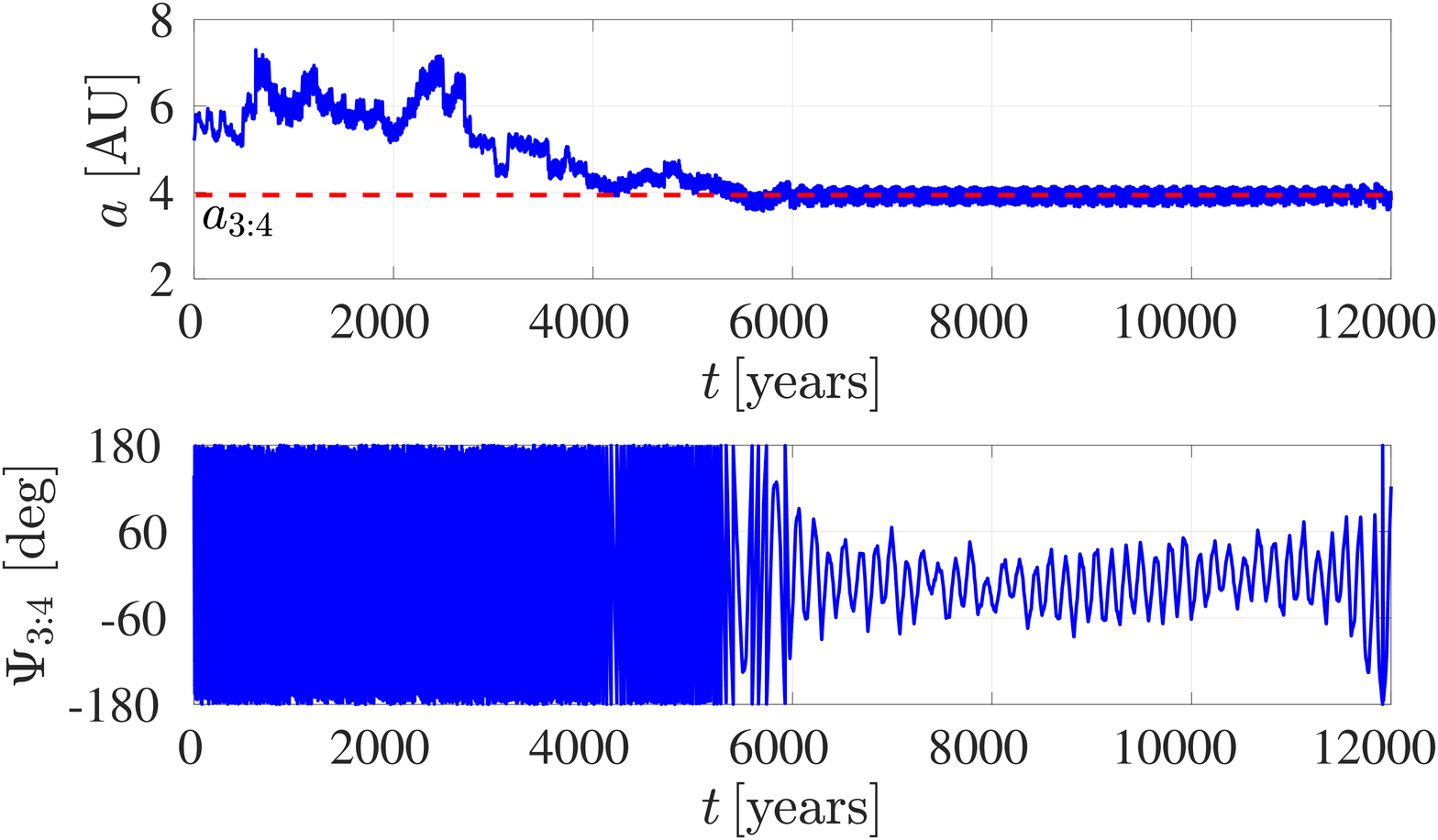}
\caption{\label{Jupiter3:4_trapped_2um}Evolution of semi-major axis and resonant argument $\Psi_\mathrm{3:4}$ for a $2 \, \mathrm{\mu m}$ particle. Here $a_{3:4}$ is for $2 \, \mathrm{\mu m}$ particles.}
\end{figure}

\begin{figure}
\centering
\noindent\includegraphics[width=9cm]{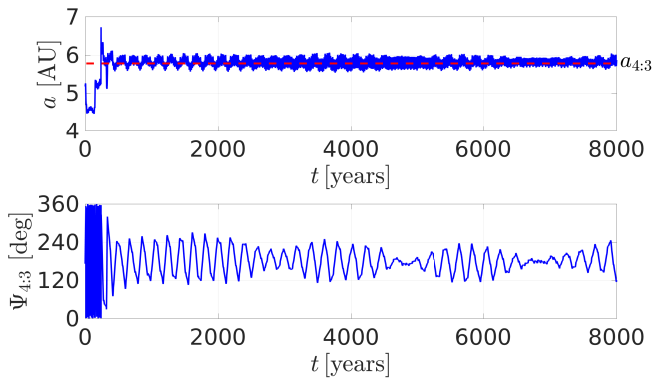}
\caption{\label{Jupiter4:3_trapped_2um}Evolution of semi-major axis and resonant argument $\Psi_\mathrm{4:3}$ for a $2 \, \mathrm{\mu m}$ particle. Here $a_{4:3}$ is for $2 \, \mathrm{\mu m}$ particles.}
\end{figure}

There are also particles that can be trapped in higher order resonances with Jupiter. As an example we show the trapping in the Jupiter (interior) 3:4 resonance (Fig.~\ref{Jupiter3:4_trapped_2um}) and (exterior) 4:3 resonance (Fig.~\ref{Jupiter4:3_trapped_2um}) for $2 \, \mathrm{\mu m}$ particles. The resonant arguments are $\Psi_\mathrm{3:4} = 3\lambda_\mathrm{eff} -4\lambda_\mathrm{Jupiter} + \varpi_\mathrm{eff}$ and $\Psi_\mathrm{4:3} = 4\lambda_\mathrm{eff} -3\lambda_\mathrm{Jupiter} - \varpi_\mathrm{eff}$. The resonant semi-major axes are calculated as $a_\mathrm{3:4} \approx 3.94 \, \mathrm{AU}$ and $a_\mathrm{4:3} \approx 5.78 \, \mathrm{AU}$. In both cases, particles are trapped for more than 5000 years. However, the trapping probabilities in the Saturn 1:1 resonance and the Jupiter higher order resonances are low. Thus, these resonances contribute little to the overall distribution of dust semi-major axis.

\section{Distributions of eccentricity $e$, longitude of pericenter $\varpi$, and inclination $i$}
\begin{figure}
\centering
\noindent\includegraphics[width=9cm]{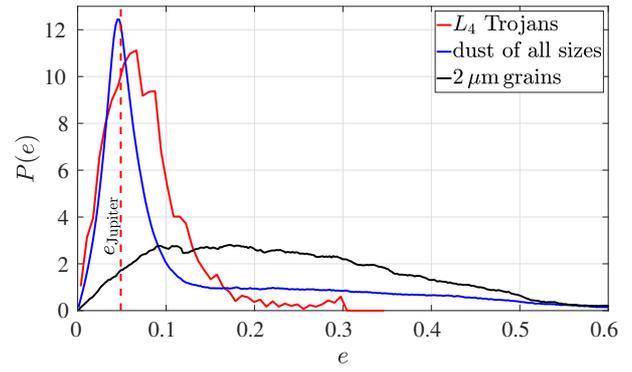}
\caption{\label{L4vsdust_e_distri}Eccentricity distribution of $L_4$ asteroids (red) and Trojan dust of all sizes (blue). The black line denotes the eccentricity distribution of $2 \, \mathrm{\mu m}$ particles.}
\end{figure}

The eccentricity distributions for the $L_4$ asteroids and Trojan dust of all sizes peak close to $e_\mathrm{Jupiter}$ ($\approx \! 0.049$), as shown in Fig.~\ref{L4vsdust_e_distri}. This is because in the elliptic restricted three-body problem, the values $a=a_\mathrm{Jupiter}$, $e=e_\mathrm{Jupiter}$, and $\varpi=\varpi_\mathrm{Jupiter} \pm 60^\circ$ are the stable equilateral solutions \citep[e.g.,][]{sandor2003symplectic, robutel2010introduction}.

The peak value of the dust eccentricity distribution is slightly smaller than $e_\mathrm{Jupiter}$. This is mainly because of the effects of P-R drag and solar wind drag. For large particles, the orbits are nearly Keplerian. Using the orbital average method, the effects of P-R drag and solar wind drag on eccentricity can be expressed as \citep{BURNS:1979wg}
\begin{equation}
\left\langle \frac{\mathrm{d}e}{\mathrm{d}t} \right\rangle _\mathrm{drag} = - (1+sw) \left(\frac{3Q_\mathrm SQ_\mathrm {pr}\mathrm{AU}^2}{4\rho_\mathrm gr_\mathrm gc^2}\right) \left(\frac{5e}{2a^2\sqrt{1-e^2}}\right) \,,
\end{equation}
where $sw$ is the ratio of solar wind drag to P-R drag, $Q_\mathrm S$ is the radiation energy flux at one AU, $Q_\mathrm{pr}$ is the solar radiation pressure efficiency, $\rho_\mathrm{g}$ is the grain density, $r_\mathrm{g}$ is the grain radius, and $c$ is the light speed. On average these two drag forces reduce the value of eccentricity for larger particles.
Large eccentricities are induced for particles with grain sizes $\leqslant 2 \, \mathrm{\mu m}$ (Fig.~\ref{L4vsdust_e_distri}). These small particles are affected by the strong solar radiation pressure and Lorentz force, which increase the probability of encounters with Jupiter, Saturn, or other planets. All these effects can excite large eccentricities.

\begin{figure}
\centering
\noindent\includegraphics[width=9cm]{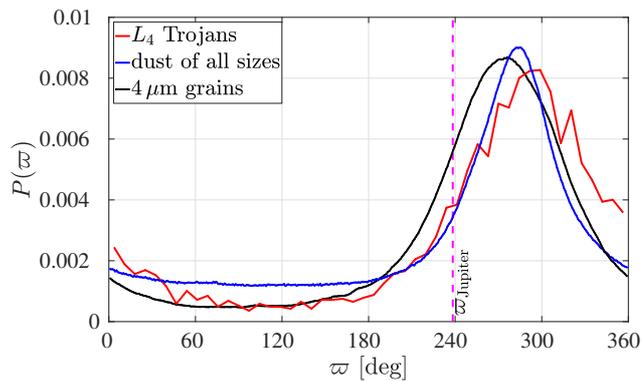}
\caption{\label{L4vsdust_lp_distri}Distribution of the longitude of pericenter for $L_4$ asteroids (red) and Trojan dust of all sizes (blue). The black line denotes the distribution of longitude of pericenter for $4 \, \mathrm{\mu m}$ particles.}
\end{figure}

We do not see obvious patterns in the distributions of $\omega$ (argument of pericenter) and $\Omega$ (longitude of ascending node). Instead, we take a look at the distribution of $\varpi = \omega + \Omega$. For the $L_4$ asteroids and their dust, the distributions of $\varpi$ are shown in Fig.~\ref{L4vsdust_lp_distri}. The peak of the Trojans' $\varpi$ distribution is about $60^\circ$ larger than $\varpi_\mathrm{Jupiter}$. This is because $\varpi=\varpi_\mathrm{Jupiter} + 60^\circ$ is the stable equilateral solution in the elliptic restricted three-body problem \citep[e.g.,][]{sandor2003symplectic, robutel2010introduction}, as mentioned above. To the knowledge of the authors, this peak with distribution of $\varpi$ for the Jupiter Trojan asteroids has not been shown in the literature to date.

Roughly, the distributions of $\varpi$ for the $L_4$ asteroids and dust have the same shape. However, we note that for Trojan dust $\Delta\varpi = \varpi_\mathrm{peak}-\varpi_\mathrm{Jupiter} < 60^\circ$. The value of $\Delta\varpi$ becomes smaller with smaller grain size, if the grains can be trapped in the Jupiter 1:1 resonance. Seen from Fig.~\ref{L4vsdust_lp_distri}, for dust of all sizes, $\Delta\varpi \approx 45^\circ$; while for 4 $\mu\mathrm{m}$ particles, $\Delta\varpi \approx 37^\circ$.

\begin{figure}
\centering
\noindent\includegraphics[width=9cm]{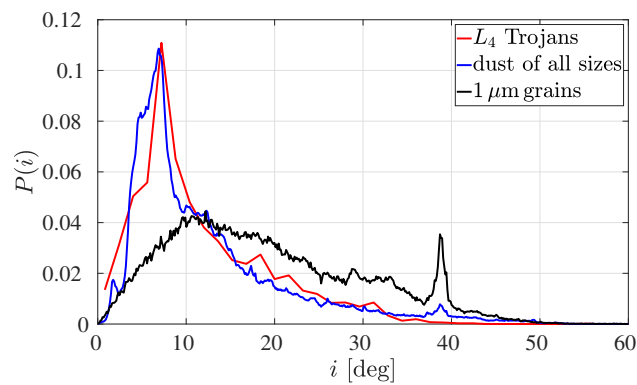}
\caption{\label{L4vsdust_i_distri}Inclination distribution of the $L_4$ asteroids (red) and Trojan dust of all sizes (blue). The black line denotes the inclination distribution of $1 \, \mathrm{\mu m}$ particles.}
\end{figure}

Compared to the distributions of $a$, $e$, and $\varpi$, the inclination distribution of dust is closer to that of the $L_4$ Trojans (Fig.~\ref{L4vsdust_i_distri}). The reason is explained as follows. The variation of the inclination can be expressed as \citep{murray1999solar}
\begin{equation} \label{eq_i_contri}
\frac{\mathrm{d}i}{\mathrm{d}t} = \frac{r\cos(\omega+f)}{na^2\sqrt{1-e^2}}N,
\end{equation}
where $r$ is the Jovicentric distance, $f$ is the true anomaly, $n$ is the mean motion, and $N$ is the normal component of the perturbation force relative to the orbital plane. Solar radiation pressure, P-R drag, and solar wind drag have no such normal component. Thus, these forces do not change orbital inclination \citep{BURNS:1979wg}. Although the Lorentz force has a normal component with respect to the dust orbital plane, for large particles the Lorentz force is weak because of its dependence on the inverse square of the grain size. Besides, the polarity of the interplanetary magnetic field rapidly changes with the 22 year period of the solar cycle \citep{gustafson1994physics,landgraf2000modeling}, so the average effect of the Lorentz force on inclination is very small \citep{fahr1995evolution,moro2003dynamical}. Also because larger particles can stay in the system for a longer time, the inclination distribution of the whole dust population is roughly consistent with that of the $L_4$ Trojans for the most part. 

\begin{figure}
\centering
\noindent\includegraphics[width=9cm]{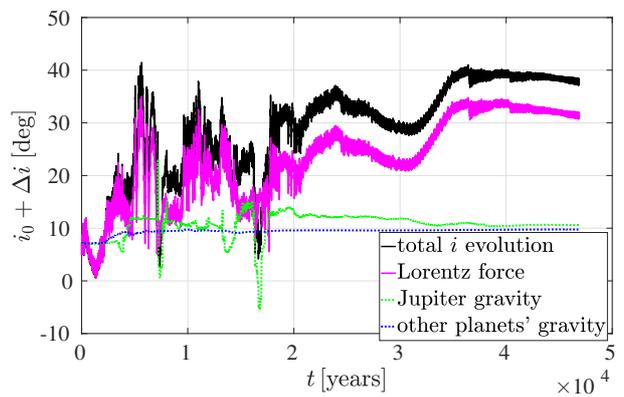}
\caption{\label{inclination_1um}Evolution of inclination starting from an initial value $i_0$. The parameter $\Delta i$ is the integral of the contribution to $\mathrm di/\mathrm dt$ due to different perturbation forces acting on a 1 $\mu \mathrm{m}$ particle.}
\end{figure}

For small particles, the magnitude of Lorentz force is relatively strong; moreover, the value of $\beta$ is relatively large, which implies that the dust orbit becomes non-Keplerian. As a result, the effect of the Lorentz force on inclination does not average out. This explains the remaining inclination difference between $L_4$ Trojans and Trojan dust (small wiggles in Fig.~\ref{L4vsdust_i_distri}). Taking $1 \, \mathrm{\mu m}$ particles for instance, they can attain high inclination (black line in Fig.~\ref{L4vsdust_i_distri}) due to the stronger Lorentz force (Fig.~\ref{inclination_1um}). 

\section{Distribution of the Tisserand parameter $T_\mathrm{P}$}
\begin{figure}
\centering
\noindent\includegraphics[width=9cm]{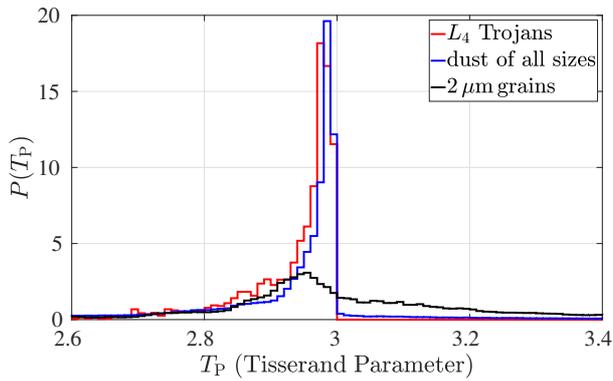}
\caption{\label{L4vsdust_tp_distri}Distribution of Tisserand parameter (relative to Jupiter) for $L_4$ asteroids (red) and Trojan dust of all sizes (blue). The black line denotes the distribution of the Tisserand parameter for $2 \, \mathrm{\mu m}$ particles.}
\end{figure}

The Tisserand parameter is a nearly conserved quantity in the circular restricted three-body problem, derived as an approximation to the Jacobi constant \citep{tisserand1896, murray1999solar}. In celestial mechanics, the Tisserand parameter is often used for orbit classification \citep{levison1996comet, levison1997kuiper}. The Tisserand parameter with respect to Jupiter is defined as
\begin{equation} \label{eq_Tisserand}
T_\mathrm{P} = \frac{a_\mathrm{Jupiter}}{a} + 2\cos i\sqrt{\frac{a}{a_\mathrm{Jupiter}}(1-e^2)} \,.
\end{equation}
Similar to the dynamical behavior of comets \citep{levison1996comet,levison1997kuiper}, typical values of the Tisserand parameter for Jupiter Trojan asteroids satisfy $2<T_\mathrm{P}<3$. From our simulations, we know that dust particles keep for most of their dynamical evolution a value of $T_\mathrm{P}$ that is close to $T_\mathrm{P}$ of their sources. As a consequence, the $L_4$ Trojans and dust particles ejected from them have similar distributions of $T_\mathrm{P}$ (Fig.~\ref{L4vsdust_tp_distri}). For the overwhelming part we have $2<T_\mathrm{P}<3$. For dust particles there is a very small portion with values of $T_\mathrm{P}>3$. This mainly attributes to particles in the size range $[1,2] \, \mathrm{\mu m}$. For these grain sizes, the lifetime is long enough for strong solar radiation pressure to increase the $T_\mathrm{P}$ values. As an example the $T_\mathrm{P}$ distribution for $2 \, \mathrm{\mu m}$ particles is also shown in Fig.~\ref{L4vsdust_tp_distri}.

For dust released from the main belt asteroids \citep{dermott2002asteroidal}, we expect $T_\mathrm{P}>3$ as their sources. Thus, determination of the orbital elements from the measured dust velocity should allow us to distinguish particles that were released from Trojan asteroids and from main belt asteroids. We note that Jupiter-family comets also have typical values of Tisserand parameters in the range $2<T_\mathrm{P}<3$ \citep{levison1996comet,levison1997kuiper}. However, for most of these comets the semi-major axes are much larger, or smaller, than those of Jupiter Trojans (5.2 AU). Besides, eccentricities of Jupiter-family comets ($>0.2$) are larger than the typical eccentricities of Jupiter Trojans. Thus, it will be possible as well to distinguish dust released from Trojan asteroids and dust from Jupiter-family comets in terms of their semi-major axes and eccentricities.

\section{Conclusions} \label{section_conclusions}
In previous work, we developed a numerical model for dust particles ejected from the Jupiter Trojans, and derived their configuration in space \citep{liu2018dust}. In this work, we compare the orbital properties of these Trojan dust particles and the Jupiter $L_4$ asteroids. We find that the peak of semi-major axis distribution for Trojan dust is smaller than that for Trojans, which is mainly due to solar radiation pressure. The distribution of the semi-major axis for grain sizes $\geqslant 4\, \mu\mathrm{m}$ is bimodal. A fraction of the particles can be trapped in the Saturn 1:1 resonance and Jupiter higher order resonances. The peak of the eccentricity distribution for Trojan dust is also smaller than that for Trojans, which is because of the action of P-R drag. For the Trojan asteroids, the peak in the longitude of pericenter distribution is about $60^\circ$ larger than Jupiter's longitude of pericenter. For Trojan dust this difference is less than $60^\circ$, and it decreases with decreasing grain size. The inclination distributions of Trojans and Trojan dust are quite similar, except that small particles can attain large inclination due to the effect of the Lorentz force. The Trojan asteroids and most Trojan dust have typical values of Tisserand parameters in the range of two to three.

\begin{acknowledgements}
This work was supported by the European Space Agency under the project Jovian Micrometeoroid Environment Model (JMEM) (contract number: 4000107249/12/NL/AF) at the University of Oulu and by the Academy of Finland. We acknowledge CSC -- IT Center for Science for the allocation of computational resources on their Taito cluster.
\end{acknowledgements}

%\bibliographystyle{aa} % style aa.bst
%\bibliography{Strings,AdditionalLit,PapersLit,lit,Ring2Galilean,trojan} % your references Yourfile.bib 

\end{document}